\documentclass[preprint,,preprintnumbers,amsmath,amssymb,superscriptaddress]{revtex4}
\usepackage[utf8]{inputenc}
\usepackage[T2A]{fontenc}
\usepackage{epsfig}
\usepackage{epstopdf}
\usepackage{amsmath,amssymb,amsthm}
\usepackage{dcolumn}
\usepackage{bm}
\usepackage{graphicx}
\usepackage{subfloat}
\usepackage{subfig}

\begin{document}

\title{General static polarizability in spherical neutral metal clusters and fullerenes \\ within Thomas-Fermi theory} 
\author{D. I. Palade}\email{p.dragos.iustin@gmail.com}
\affiliation{National Institute of Laser, Plasma and Radiation Physics,
PO Box MG 36, RO-077125 Magurele, Bucharest, Romania}
\author{V.Baran}\email{virbaran@yahoo.com}
\affiliation{Faculty of Physics, Bucharest}


\begin{abstract} 
We study the static linear response in spherical Thomas-Fermi systems deriving a simple differential equation for general multipolar moments and associated polarizabilities. We test the equation on sodium clusters between $20$ and $100$ atoms and on fullerenes between $C_{60}$ and $C_{240}$ and propose it for general Thomas-Fermi systems. Our simple method provides results which deviates from experimental data with less then $10\%$.
\end{abstract}

\maketitle

\section{Introduction}
The problem of linear response to an external field is a crucial problem in many-body physics and it has been intesively studied through different techniques. For the static case, the problem simplifies itself to static response functions as polarizability and magnetic susceptibility, crucial quantities in the description of any classical or quantum system. 

In the range of mesoscopic systems as metallic clusters are, various classical methods have been applied successfully during the first part of the 20th century, especially Mie's theory for electromagnetic scattering and electrostatic modeling for the polarizability. Nonetheless, going down on the length scale, the classical models start to fail such that in the atomic, molecular and mesoscopic domains, the predicted results are no longer consistent with the experimental data.

Semi-classical or fully quantum models are appropriate to describe the observed features. Methods as Hartree-Fock theory, RPA, Density Functional Theory, etc., in general, mean field theories, give very good results regarding the stationary (or dynamic) properties of such atomic systems. Nonetheless, they have a single flaw: the computational efforts are huge and may not worth to use such complex methods to derive simple quantities as polarizability, especially when the rigor of result is not necessary.

On the other hand, there are large classes of systems, like metallic clusters, in which some special properties are exhibited. For example, in alkali elements it is well known the presence of a quasi-free electron on the last unclosed shell which is close to the model of free homogeneous electron gas (HEG). This type of thinking it is used in solid-state physics and can be exploited in the atomic domain through the Thomas-Fermi model \cite{lieb1977thomas,thomas1927calculation,drake2006springer} which approximates the problematic term of kinetic energy in the electron system with the local form of HEG and simplifies the treatment. Extended versions of it can employ also additional terms for the exchange-correlation potential (from Kohn-Sham potential)\cite{drake2006springer} or supplementary gradient correction from the Weizsacker term \cite{weizsacker1935theorie}.

Even if such an approximation provides a very simple way to obtain the electron density, the systems in which we apply it must be carefully chosen, since usual errors can be around $10-20\%$, or larger, for different observables and due to "no-binding" Teller's theorem \cite{jahn1937stability} in molecules appear the phenomenon of instability. Even though the extended versions provide better results, we will refer in the present work just on the simple, original form of the theory since our main goal is to achieve quantitative description with minimum of computational effort. That is why, the best suited cases are those with a large volume extension, metallic or transition character and in which, we are not interested in fine details of the density profile or single particle (pseudo)wave functions, shell effects, excitation energies, etc.

For all those reasons, we shall focus next on the Thomas-Fermi theory and on modeling the linear response to a static external potential in its most general form. The theoretical results will be tested on various medium-sized sodium clusters and on the famous $C_{60}$ fullerene and $C_{240}$ and show that the method provides resonable results.

\section{Formal background}
\subsection{Metallic clusters and the jellium model}

Clusters are by definition, mesoscopic systems formed from $3-10^7$ \cite{reinhard2008introduction} atoms of various elements. The theoretical methods of investigation are going from molecular to bulk (solid-state) domain, both quantum and classical approaches.

As we have mentioned before, the clusters formed from metallic elements have the property that the electrons from the unclosed shell are loosely bounded and their behavior is close to HEG one. Also, we treat the problem in the Born-Oppenheimer approximation. This means that their dynamics and de-localization  are high enough to consider that they $see$ the ionic background in a averaged manner. By $ionic$ $background$ we understand the system obtained from the coupling between charged nuclei and the $core$ electrons which determine a net positive charge.

Our working framework is the so called $jellium$ $model$ in which the ionic background is approximated to a homogeneous positive charge distribution while the free electrons will also have an almost constant density in the bulk region. In clusters the jellium model it is also applied and the ionic background determines a smooth Coulomb potential with the appropriate symmetries. Once the Coulomb potential is generated it may be included as an input in different theoretical approaches (Hartree-Fock, DFT, etc.) to derive the electron density.  
The self-consistent jellium model proved to be a very appropriate method predicting quantitative results in good agreement with the experimental data \cite{brack1993physics}, \cite{de1993physics}. The simplest geometry possible is that of a sphere, but can be somehow a troublemaker in numerical simulations since it has a discontinuity at the edge. A more refined model is the so called $soft$ $jellium$ which works with a Wood-Saxon radial profile coupled with any possible angular dependence, giving access to any possible geometry of the ionic part. The jellium density is written in its most general form as:

\begin{equation}\label{jellium}
\rho_{jel}(r,\theta,\phi)=\frac{3}{4\pi r_s^3}[1+exp(\frac{|r|-R(\theta,\phi)}{\sigma_{jel}})]^{-1}
\end{equation} 

with $R(\theta,\phi)=R_0(1+\sum_{l,m}\alpha_{lm} Y_{lm}(\theta,\phi))$. The Weitzecker-Seitzs radius $r_s$ is a parameter of the bulk domain interpreted as the volume occupied by a single atom while $Z$ is the difference between the numbers of protons and the number of bound electrons and so $Z|e|$ is the charge of jellium. Therefore, the jellium density satisfies the condition: $\int\rho_{jel}dr^3=Z$. The limiting case of spherical sharp distribution is obtained when $\sigma_{jel}\to 0$.

\subsection{Thomas-Fermi model}

Thomas-Fermi model (TF) was derived independently by L. Thomas and E. Fermi in 1927, soon after Schroedinger equation, 1926. The basic approximation of the model is to treat the electron density distribution in the atom using a local approximation for the kinetic term i.e. the HEG approximation: $E_{kin}=\frac{3\gamma}{5}\int_{\mathbb{R}^3}\rho^{5/3}(\vec{r})\mathrm{d}r^3$ , with $\gamma=(3\pi^2)^{2/3}\hbar^2/2m$. If $V(\vec{r})$ is the external potential , the total energy density functional can be written:

\begin{equation}
	E[\rho(\vec{r})]=\frac{3\gamma}{5}\int_{\mathbb{R}^3}\rho^{5/3}(\vec{r})\mathrm{d}r^3+e\int_{\mathbb{R}^3}\rho(\vec{r})V(\vec{r})\mathrm{d}r^3+\frac{1}{2}\frac{e^2}{4\pi\varepsilon_0}\int_{\mathbb{R}^3}\frac{\rho(\vec{r})\rho(\vec{r'})}{|\vec{r}-\vec{r'}|}\mathrm{d}r'^3
\mathrm{d}r^3\end{equation}

The ground-state energy and the corresponding density distribution are obtained within a Ritz variational principle, searching for the minimum of this quantity:

\begin{equation}
E^{TF}=min\{E[\rho(\vec{r})]|\ \rho \in \mathfrak{L}^{5/3}(\mathbb{R}^3),\ \int_{\mathbb{R}^3}\rho(\vec{r})\mathrm{d}r^3=N,\ \rho(\vec{r}) \geq 0\}
\end{equation}

The condition $ \rho \in \mathfrak{L}^{5/3}(\mathbb{R}^3)$ refers to the fact that density is a $5/3$ integrable function over $\mathbb{R}^3$ and so, is a condition for finite kinetic energy to emerge. The constrain associated with the condition $\int_{\mathbb{R}^3}\rho(\vec{r})\mathrm{d}r^3=N$ is introduced through Euler-Lagrange multiplier technique and solving the variational problem, the resulting Thomas-Fermi equation is be derived:

\begin{equation}
\gamma\rho^{2/3}=max[0,\Phi-\mu]
\end{equation}

In this equation, $\Phi(\vec{r})$, $\Phi:\mathbb{R}^3\to\mathbb{R}$ is the energetic Coulomb potential while $\mu$ is the chemical potential:

\begin{equation}
\Phi(\vec{r})=|e|V(\vec{r})-\frac{e^2}{4\pi\varepsilon_0}\int_{\mathbb{R}^3}\frac{\rho(\vec{r'})}{|\vec{r}-\vec{r'}|}\mathrm{d}r'^3
\end{equation}

\begin{equation}
\mu=-\frac{\partial E^{TF}}{\partial N}
\end{equation}

In turn, the Coulomb potential is the generated by the charge distribution and is the solution of the Poisson's equation  $\Delta \Phi(\vec{r})=|e|\Delta V(\vec{r})+e^2\rho(\vec{r})/\varepsilon_0$. In the case of neutral electric systems, the chemical potential is null \cite{lieb1977thomas}, but this feature is maintained only within the Thomas-Fermi method. The additional terms, as Dirac or Wietzacker contributions, break this property. 
In the jellium approximation $V(\vec{r})$ is induced by $\rho_{jel}$ and the Thomas-Fermi becomes in differential form:

\begin{equation}
\Delta\Phi(\vec{r})=e^2/\varepsilon_0(\gamma^{-3/2}\Phi^{3/2}(\vec{r})-\rho_{jel}(\vec{r}))
\end{equation}

\section{Theory}

\subsection{Perturbation theory and density changes}

In the absence of external interactions, for the ground-state, the TF equation leads to $\gamma \rho_0^{2/3}(r)=\Phi_0(r)$ with $\rho_0$ the ground state density of the free electrons. We shall study in the following the static linear response in TF approximation considering a time-independent potential of an arbitrary form. If the coupling strength to the free electrons is $\lambda$:

\begin{equation}
v(\vec{r})=\lambda \sum \frac{v_{lm}(r)}{r}Y_{lm}(\theta,\phi)
\label{potexpansion}
\end{equation}

And the TF equation for the new stationary state, becomes:

\begin{equation}
\gamma \rho^{2/3}(\vec{r})=\Phi(\vec{r})
\end{equation}

Then the interaction energy density is $\rho v(\vec{r})$. This perturbation will induce a spatial change of the charge which can be treated in a power expansion in the coupling strength $\lambda$: $\rho(\vec{r})=\rho_0(r)+\lambda\rho^1(\vec{r})+\lambda^2\rho^2(\vec{r})+ ...$.

If we resume to first order term $(\lambda\ll e^2/(4\pi\varepsilon r_0))$,  then $\rho^1$ should satisfy
$\int_{\mathbb{R}^3}\rho^1(\vec{r})\mathrm{d}r^3=0$ since  $\int_{\mathbb{R}^3}\rho_0(\vec{r})\mathrm{d}r^3=\int_{\mathbb{R}^3}\rho(\vec{r})\mathrm{d}r^3=Z$.

In the presence of the external potential the initial spherical symmetry is broken and the densities varify the properties:

\begin{subequations}
\begin{align}
       \rho : \mathbb{R}^3\to\mathbb{R}_+ \\
       \rho_0 : \mathbb{R}^3\to\mathbb{R}_+ \\
       \rho^1 : \mathbb{R}^3\to\mathbb{R}
\end{align}
\end{subequations}

The linearized kinetic density energy term and the potential $\Phi$ are:

\begin{equation}
\gamma \rho^{2/3}(\vec{r})=\gamma \rho_0^{2/3}(r)+\lambda\frac{2\gamma}{3\rho_0^{1/3}(r)} \rho^1(\vec{r})
\end{equation}

\begin{equation}
\Phi(\vec{r})=\Phi_0(\vec{r})+v(\vec{r})-\lambda\int\frac{\rho^1(\vec{r})}{|\vec{r}-\vec{r'}|}dr'^3
\end{equation}

Then the TF equation for the perturbed part becomes:

\begin{equation}
\lambda\frac{2\gamma}{3\rho_0^{1/3}(\vec{r})} \rho^1(\vec{r})=v(\vec{r})-\lambda\int\frac{\rho^1(\vec{r})}{|\vec{r}-\vec{r'}|}dr'^3
\label{eq:tfro1}
\end{equation}

Working in spherical coordinates we can consider the following expansion of $\rho^1$

\begin{equation}
\rho^1(\vec{r})=\rho_0^{1/3}(r)\sum_{lm}Y_{lm}(\theta,\phi) \frac{u_{lm}(r)}{r}
\label{densexpansion}
\end{equation}

Using the expansions for potential $\eqref{potexpansion}$ and for density $\eqref{densexpansion}$ in equation $\eqref{eq:tfro1}$ we deduce that $u_{lm}(r)$ functions will satisfy the radial equation:
  
\begin{equation}
\frac{d^2u_{lm}(r)}{dr^2}- u_{lm}(r)(\frac{l(l+1)}{r^2}+\frac{6\pi}{\gamma}\rho_0^{1/3}(r))=\frac{d^2v_{lm}(r)}{dr^2}- \frac{l(l+1)}{r^2}v_{lm}(r)
\label{eq:ultimate}
\end{equation}

Our equation is an approximate version of the equation (23) deduced in \cite{serra1989static} in the absence of Weizsacker, exchange and correlation terms.
This fact can be seen by setting the $beta$ factor to $0$ and excluding exchange-correlation effects. Nonetheless, their derivation is done on the basis of variational method, while our is not. We stress in this paper the fact that the complexity of the equation proposed in \cite{serra1989static} is much more involved being an integro-differential equation and the presence of those supplementary terms is not necessary for semi-quantitative results. In fact, the results have the same level of accuracy with ours and since the present equation is pure differential we can take advantage of its simplicity in order to study the main effects which contribute to polarizability in metal clusters. Moreover, while in the reference, the Weiszacker correction is used with different constants and is considered to be the essential reason for which the differential equation is derived, our derivation has no such condition.

\subsection{Boundary condition and general polarizability}

Concerning the boundary conditions, in the origin, in order to have finite $\rho^1(0)$ we require that $\lim\limits_{r\to 0}u_{lm}(r)/r$  to be finite. Concerning the behavior at infinity we shall ask for the coefficient $u_{lm}(r)$ to follow the behavior of the perturbation i.e. $u_{lm}(r)=3/(2\gamma)v_{lm}$ when $r\to\infty$. From TF eq $\eqref{eq:tfro1}$, the asymptotic behavior of $u_{lm}$ is:

\begin{equation}\label{eq:cond1}
u_{lm}(r)\to \frac{3}{2\gamma}(v_{lm}-\frac{4\pi}{2l+1}\frac{q_{lm}}{r^l})
\end{equation}

Here, the $q_{lm}$ term is the multipole moment associated with the induced charge:

\begin{equation}
q_{lm}=\int_0^\infty u_{lm}(r')\rho_0^{1/3}(r)r'^{l+1}dr'
\end{equation}

From numerical point of view we solve the equation $\eqref{eq:tfro1}$ with the associated boundary conditions as it follows: first we guess the term $q_{lm}$ (considering the particular system to be studied, the magnitude can be easily guessed) and solve the equation in such a way that the solution satisfies the asymptotic behavior mentioned above for the selected value of $q_{lm}$. With the solution constructed in these way  we find a new value of $q_{lm}$ and repeat the procedure until we reach convergence condition of the solution.

Even though can seem to be a long road to the convergence, in practice, this is reached within 10 iterations.

Again, in comparison with \cite{serra1989static} we have different asymptotic boundary condition, simply from the form of our equation and the meaning of the unknown. The entire method of iteration for finding the polarizability as a parameter of asymptotic behavior is original, at the best of our knowledge and essential for the results. The above mentioned reference does not discuss such matters.

\subsection{Dipole case}

In this section we shall apply the method described above to the specific case of dipolar response. The applied field is $v(\vec{r})=rcos\theta$ and consequently $v_{10}=r^2$. The equation $\eqref{eq:ultimate}$ becomes:  

\begin{equation}
\frac{d^2u_{10}(r)}{dr^2}- u_{10}(r)(\frac{2}{r^2}+\frac{6\pi}{\gamma}\rho_0^{1/3}(r))=0
\end{equation}

Of course, this can and must be solved numerically, but there are some specific cases with analytic solutions which can be helpful for the boundary conditions. 

For $\rho_0(r)=0$, which usually describes the regions with large $r$ where the density must be null we have the solution which respects the boundary condition (19) for $q_{10}$, $u_{10}(r)=Ar^2+\frac{B}{r}$. In the more general case of $\rho_0(r)=const$ we have a more elaborate solution ($k^2=6\pi\rho_0^{1/3}/\gamma$):

\begin{equation}\label{eq:cond2}
u_{10}(r)\propto(e^{kr}(\frac{1}{2k^3r}-\frac{1}{2k^2})-(e^{-kr}(\frac{1}{2k^3r}+\frac{1}{2k^2}))
\end{equation}

\section{Results and discussion}
\subsection{Na clusters}

Sodium clusters represent a textbook metal cluster due to the nature of the element which has a single electron on the last shell, namely the $Na$ with the atomic number $Z=11$ and the electronic configuration $1s^22s^22p^63s^1$. This element has been taken into account in this work, due to their close to spherical symmetry for medium sized clusters and due to the strong metallic character. The classical electromagnetism provides, in the frame of small metal sphere model, a polarizability connected with the radius
by:

\begin{equation}
\alpha_{classic}=R^3
\label{clasic}
\end{equation}

Experimental data reveals higher poralizabilities for all $Na$ clusters, only in the high radius limit, the classical value is reached.

\begin{figure}[!htbp]
\subfloat{\includegraphics[width = 54mm]{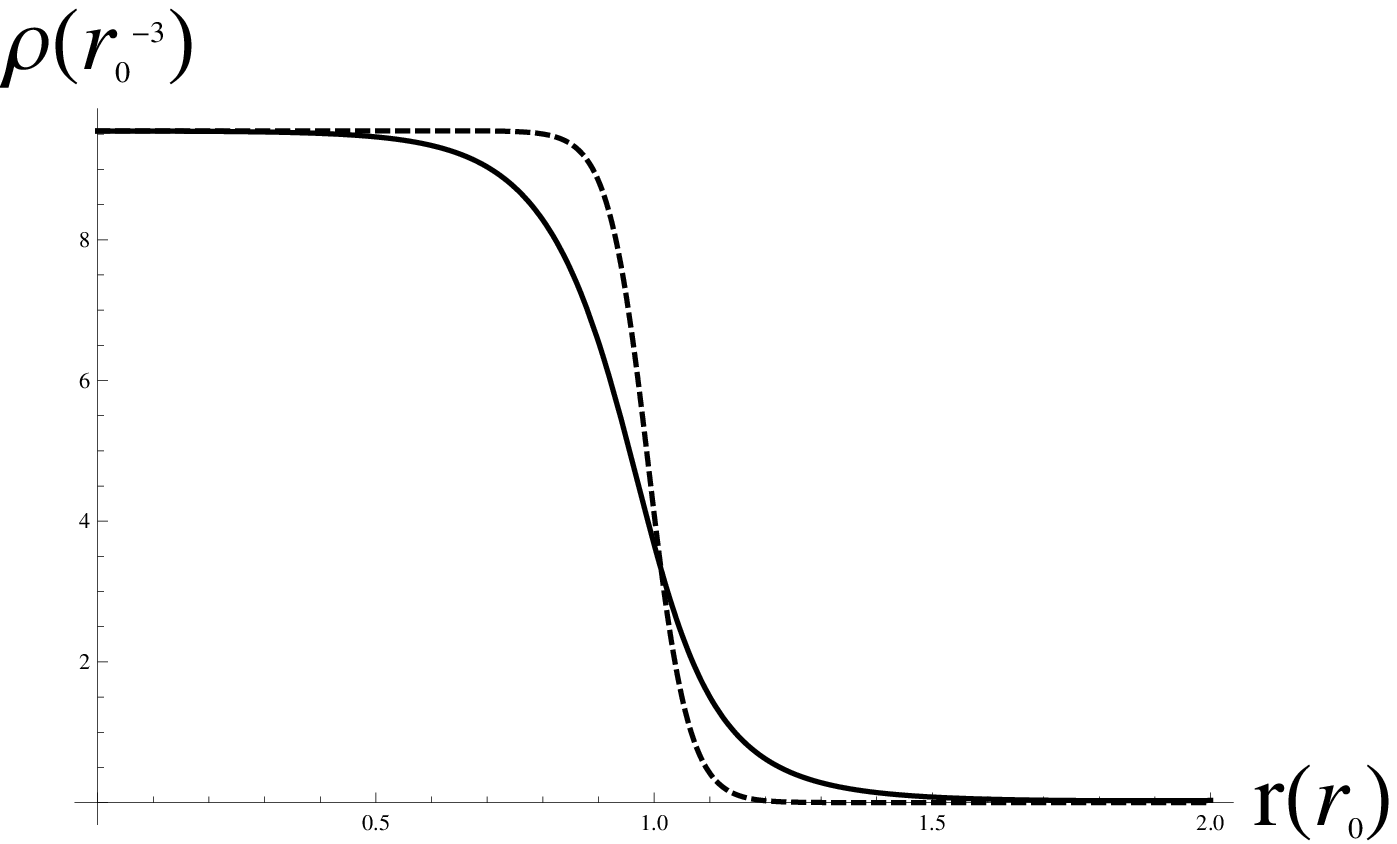}}\hspace{2mm} 
\subfloat{\includegraphics[width = 54mm]{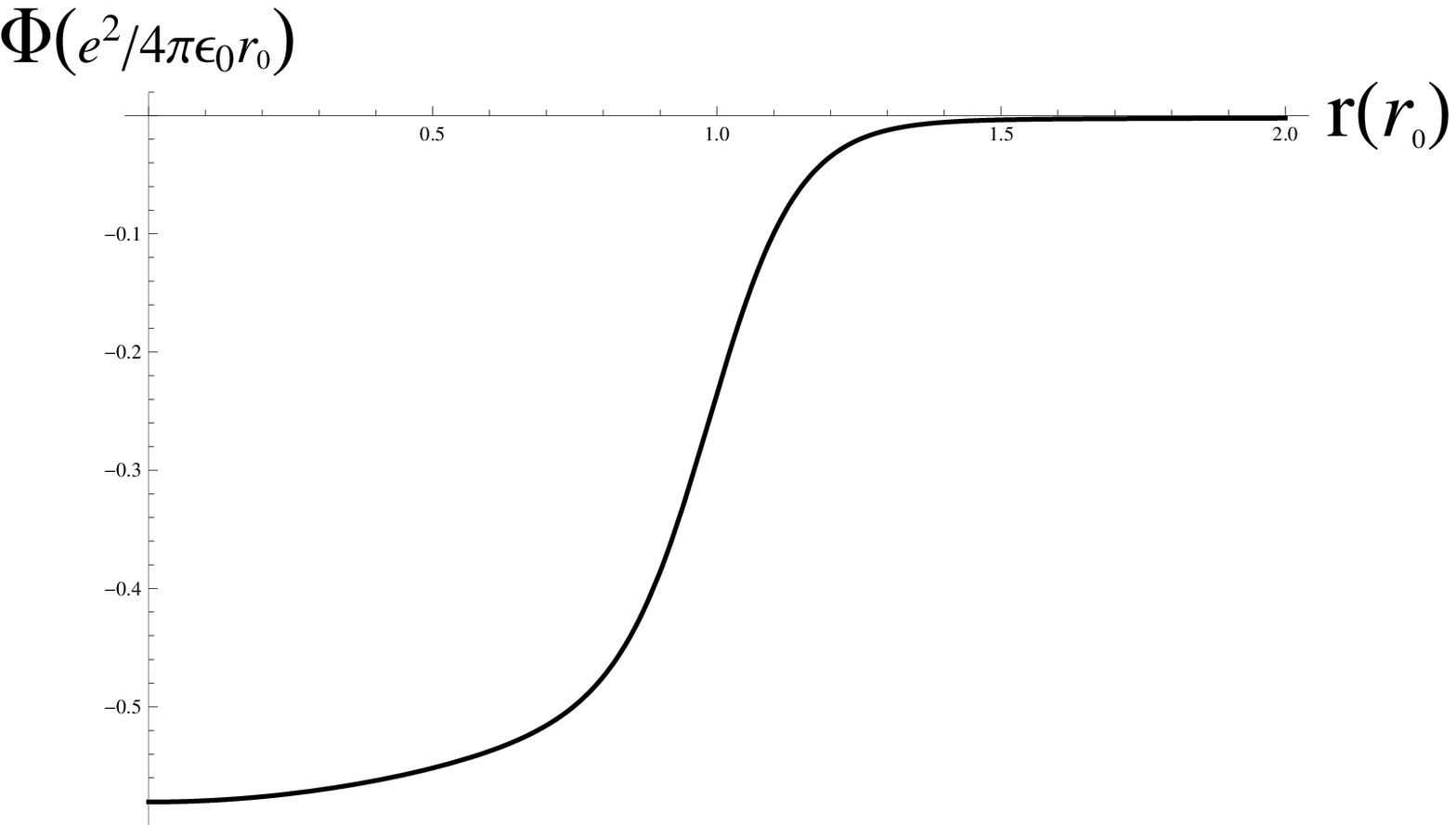}}\\
\subfloat{\includegraphics[width = 54mm]{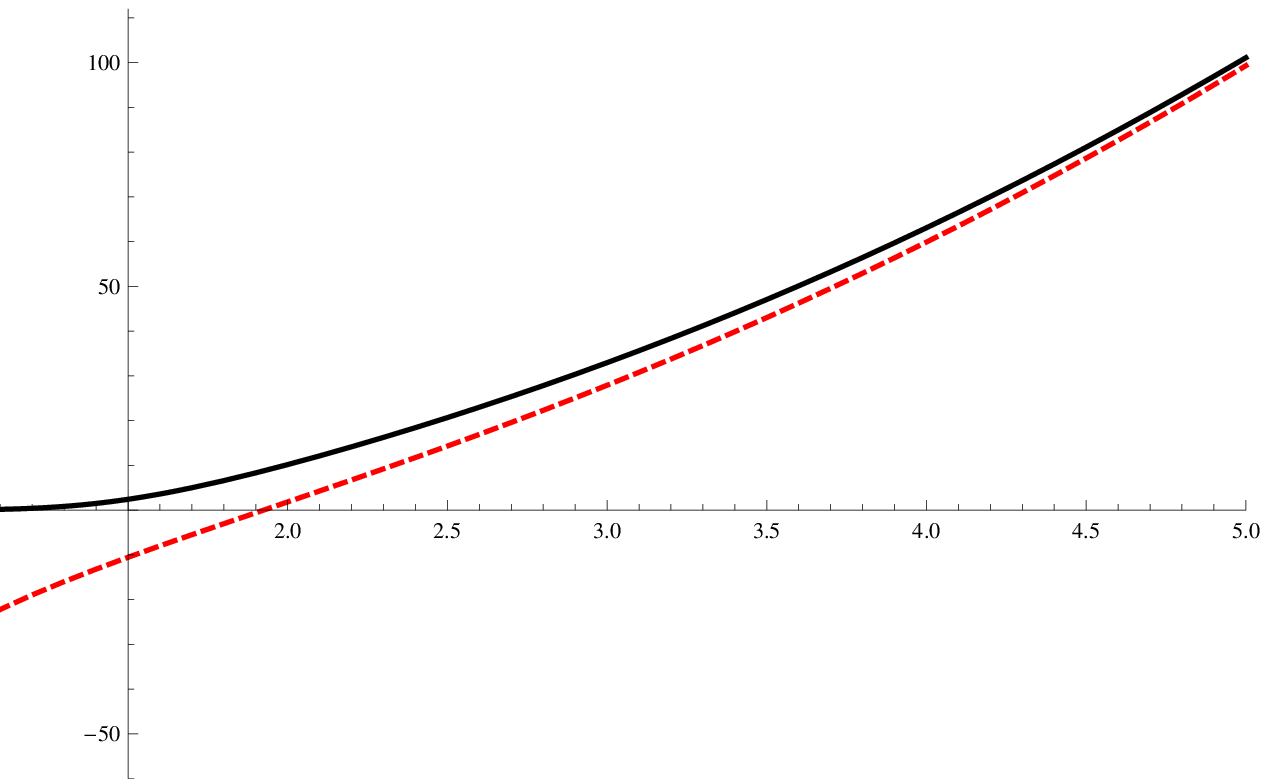}} \hspace{3mm}
\subfloat{\includegraphics[width = 60mm]{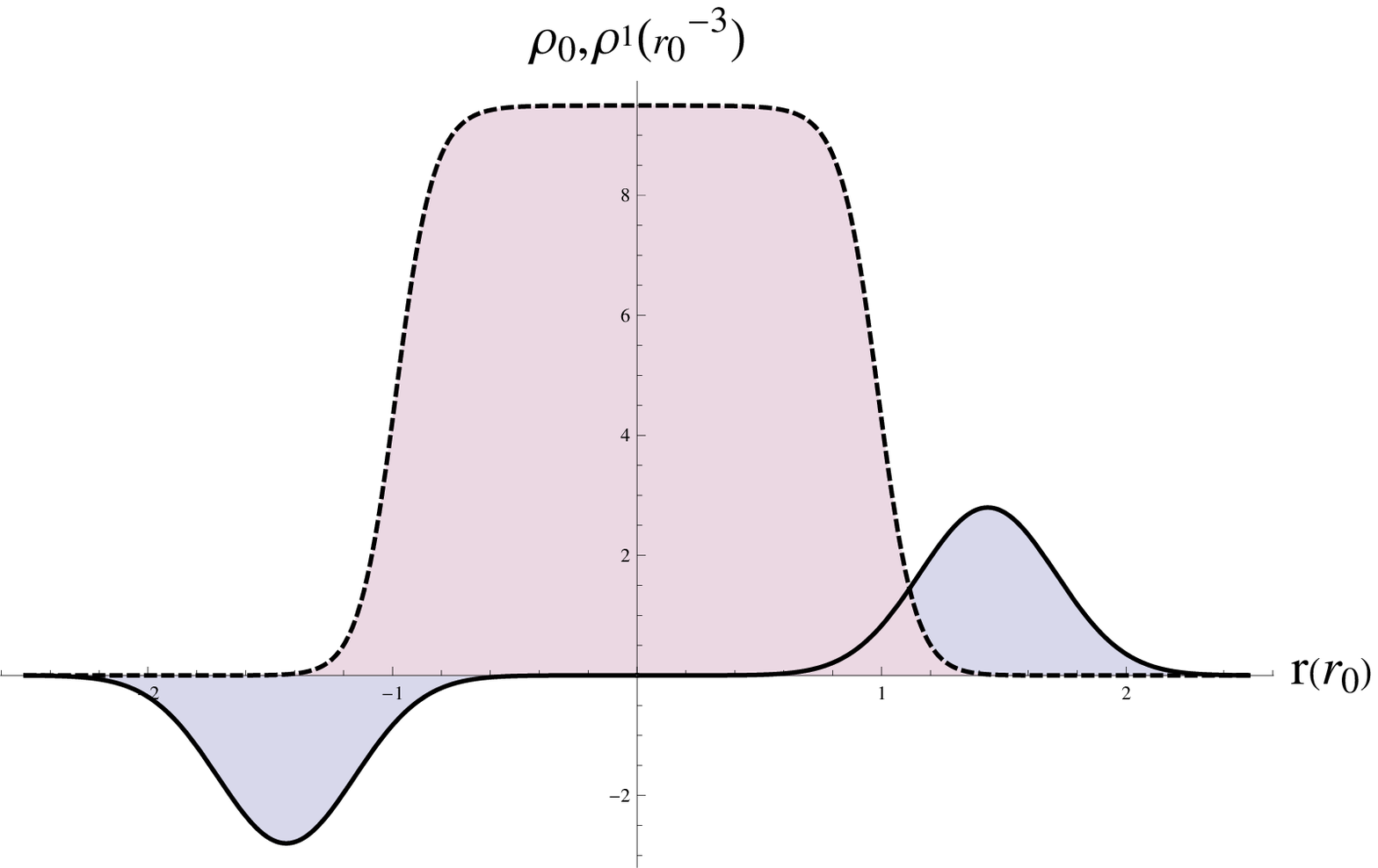}}
\caption{\footnotesize{a)Jellium density (Dashed line) and electron density (continuous line) $\rho$ for the $Na$ cluster with $N=40$;
b) Ground-state electrostatic potential $\Phi_0$ for the $Na$ cluster with $N=40$;
c)Fitting the solution $u_{10}(r)$ (continuous) to the asymptotic function (red,dashed) at large distances for the $Na$ cluster with $N=40$;
d) Ground-state electron density (dashed) and induced charge density (continuous) on the $Oz$ direction.( The induced charge was intentionally raised up in order to have a visible effect. in reality, its effect is much smaller than ground state so no negative density region can arise)}} 
\label{figNa}
\end{figure}

In our calculations, different clusters were taken into account as having spherical symmetry and a constant density of atoms. The electrons on the first two atomic levels were considered as core electrons and so, the jellium model reduces to a sphere of a radius connected to the number of atoms through the  Weitzecker-Seitzs radius. The positive jellium charge of $N|e|$ distributed by a Wood-Saxon profile like in $\eqref{jellium}$ (but no angular dependence), with a sharp fall of density around the radius of the cluster ($\sigma\simeq 0.8a_0$ usually used \cite{reinhard2008introduction}) and Seitz radius $r_0=3.93 a_0$, atomic units of length, see Fig.~\ref{figNa}a), for the case $N=40$. Other parametrization of smaller $\sigma$ have been explored, but due to sensitivity of the method far from the center of the cluster, this parametrizations give worse results. 

Taking into account the spherical symmetry, the TF equation reduces to radial differential equation:

\begin{equation}
\frac{d^2}{dr^2}(\frac{\Phi_0(r)}{r})=4\pi r ((\frac{\Phi_0(r)}{r\gamma})^{3/2}-\rho_{jel}(r))
\label{TFNa}
\end{equation}

For the same number of atoms, in Fig ~\ref{figNa}a) (continuous line) is plotted the electron density $\rho$ as obtained from above equation $\eqref{TFNa}$.
In practice, clusters with the number of atoms between $20$ and $100$ were investigated, but in Fig. ~\ref{figNa} just a generic plot of the jellium density and the electron density is presented in units of $1/r_0^3$, with $r_0$ the radius of the cluster.

This distribution manifests a tail beyond the jellium volume associated with the quantum behavior of the electrons. The ground state electrostatic potential $\Phi_0$ is plotted in Fig. ~\ref{figNa}b), while the radial dependence of the induced charge density is represented in Fig ~\ref{figNa}c) (blue-filled). In figure Fig. ~\ref{figNa}c) we're drawn for comparison the asymptotic function as described by $\eqref{eq:cond1}$ and actual solution.

\begin{figure}[!htbp]
\centering
\includegraphics[width=0.7\linewidth]{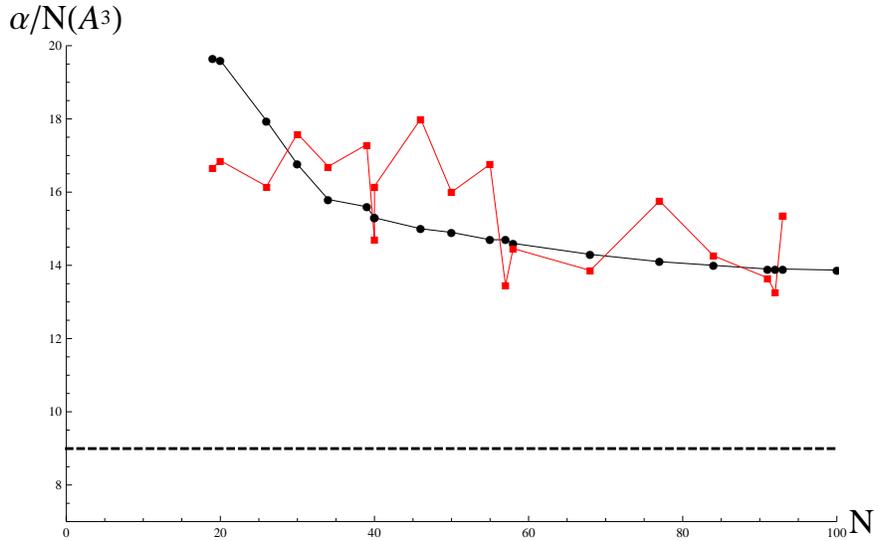}
\caption{Theoretical vs Experimental polarizability in $Na$ \cite{tikhonov2001measurement}}
\label{experNa}
\end{figure}

The proposed equation $\eqref{eq:ultimate}$ has been used in the dipolar case with the analytic limits $\eqref{eq:cond1}$,$\eqref{eq:cond2}$ and the polarizability was obtain in a quite good agreement with the experimental results\cite{tikhonov2001measurement}. The results can be seen and compared with reference \cite{tikhonov2001measurement} in tabel ~\ref{tabel1} and in ~\ref{experNa} where the quantity $\alpha/N$ as a function of $N$ is plotted. The solid horizontal line is associated with the classic solution.

\begin{table}[!htbp]
\caption{ Static polarizability of $Na$ clusters with $20<N<100$} 
\centering 
\begin{tabular}{c c c} 
\hline\hline 
N & Exp. & Result \\ [0.5ex] 
\hline 
19 & 16.66 & 19.66\\
20 & 16.86 & 19.6\\
26 & 16.16 & 17.944\\
30 & 17.6 & 16.76\\
34 & 16.7 & 15.8\\
39 & 17.3 & 15.6\\
40 & 14.7 & 15.3\\
40 & 16.16 & 15.3\\
46 & 18. & 15.\\
50 & 16. & 14.9\\
55 & 16.76 & 14.7\\
57 & 13.46 & 14.7\\
58 & 14.46 & 14.6\\
68 & 13.86 & 14.3\\
77 & 15.76 & 14.1\\
84 & 14.26 & 14.\\
91 & 13.66 & 13.9\\
92 & 13.26 & 13.9\\
93 & 15.36 & 13.9\\
93 & 15.36 & 13.87\\[1ex] 
\hline 
\end{tabular}
\label{tabel1} 
\end{table}

For low numbers of atoms, the equation fail to describe quantitatively the polarizability due to the fact that the jellium model and the spherical shape of the cluster are no longer realistic approximations while in the range $N=40$, $N=100$ we can see an error bellow $10\%$ (plotted in ~\ref{experNa}), the only deviations being a consequence of shell effects. Also, the tendency of decreasing to a constant (bulk) value with the number of atoms involved, explained as a classical limit of our semi-classical treatment of the electron system.

\begin{figure}[!htbp]
\centering
\includegraphics[width=0.6\linewidth]{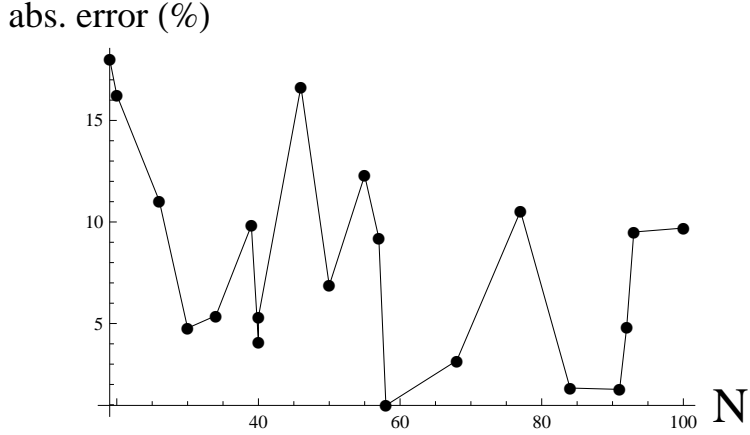}
\caption{Relative error of obtained polarizability}
\label{fig:errorNa}
\end{figure}

From numerical solution obtained with the above method, we have observed that it is possible an empirical parametrization of the approximative $u_{10}(r)$ solution in the general case as:

\begin{equation}
 u_{10}(r)=\frac{3}{2\gamma}(r^2-\frac{4\pi}{3r}(1-e^{-\beta r}))
\end{equation}

Which allows us to formulate a final empirical sum rule-like expression for static dipole polarizability. The potential of this observation is that links the linear response only to the ground state properties of the system as in the usual moments of the response function:

\begin{equation}
 \alpha_{10}(r)=r_0^3\frac{\int\limits_{0}^{\infty}\rho_0^{1/3}(r)r^4dr}{ \frac{2\gamma}{3}+\frac{4\pi}{3}\int\limits_{0}^{\infty}\rho_0^{1/3}(r)(1-e^{-(0.219-25.52/N^2) r})rdr}
\end{equation}

 We make comparison with the well known sum rule for static dipole polarizability [\cite{de1993physics}] exhibited by spherical metal clusters in which the main contribution is given through the so called $spilled-out$ electron which are considered outside the jellium region and appear proportional to $\delta$ in the approximation : $\alpha\simeq r_0^3(1+\delta)$ \cite{snider1983density}.

\subsection{$C_{60}$ fullerene}

The Buckminster fullerene represents one of the most studied molecule in the last decades due to its high symmetry, special features, high stability, etc. Consequently, the polarizability has been studied \cite{gueorguiev2004quantum} in many models, the best theoretical results being obtained in the frame of DFT-LCAO  while other theories obtained large errors in respect with the experimental data.

\begin{figure}[h]
\begin{center}
\includegraphics[width=30mm]{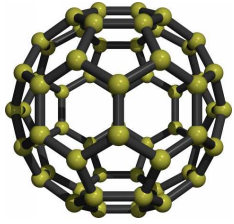}
\end{center}
\caption{\small{Geometry of C60 fullerene}}
\end{figure}

Essentially, fullerene is a carbon molecule with the atoms placed on a structure similar with that of a soccer ball. Due to the fact that carbon is not a genuine metal, one could argue that to study it along with true metallic clusters like sodium, it is a bad mistake. Nonetheless, even if from electronic structure and band gap point of view our approach can not be justified, it is a known fact that the optical spectra from fullerene exhibit a large, well localized plasmon. Further, this plasmon it is explained as being a surface plasmon \cite{scully2005photoexcitation} and so, it can be reasonably concluded that in the dynamic regime, the electrons from fullerene behave close to the ones from a metal. For this reason we have applied Thomas-Fermi equation for the semi-delocalized electrons and obtain a good description for polarizability, Mie's plasmon centroid and the density distribution for the ground-state.

We have used as jellium model, a gaussian distribution centered on the radius on which the carbon nuclei are placed but with a small width, described by the equation $\rho_{jel}(r)\propto exp(-\sigma(r-r_0)^2)$. Regarding the charge, our jellium model contain the core electrons from the $1s^22s^2$ while the other 240 electrons from $2p^2$ are considered quasi-free and taken into account in the TF calculations. The results are quite sensitive to the width ($\Delta=$ full width at half maximum) of the jellium gaussian distribution and for that reason we have performed our calculations with different values for this quantity between $0.01 \mathring{A}$ and $0.6 \mathring{A}$. This impediment is hard to be avoided since the physical meaning of this width is the radius of the core electrons in which they can be accounted as part of jellium, but in an approximative physical way, our values cover the usual atomic values for this feature and provide a set of very close values to the experimental data. 

Also, another approach to the jellium model was considered the spherical homogeneous shell (discussed in almost all papers using jellium model in $C_{60}$ investigations \cite{rudel2002imaging}\cite{puska1993photoabsorption}\cite{weaver1991electronic}) centered on the mean radius of $r_0=3.54 \mathring{A}$ and with a width of $1.5 \mathring{A}$\cite{rudel2002imaging}.

\begin{figure}[h]
\subfloat{\includegraphics[width = 54mm]{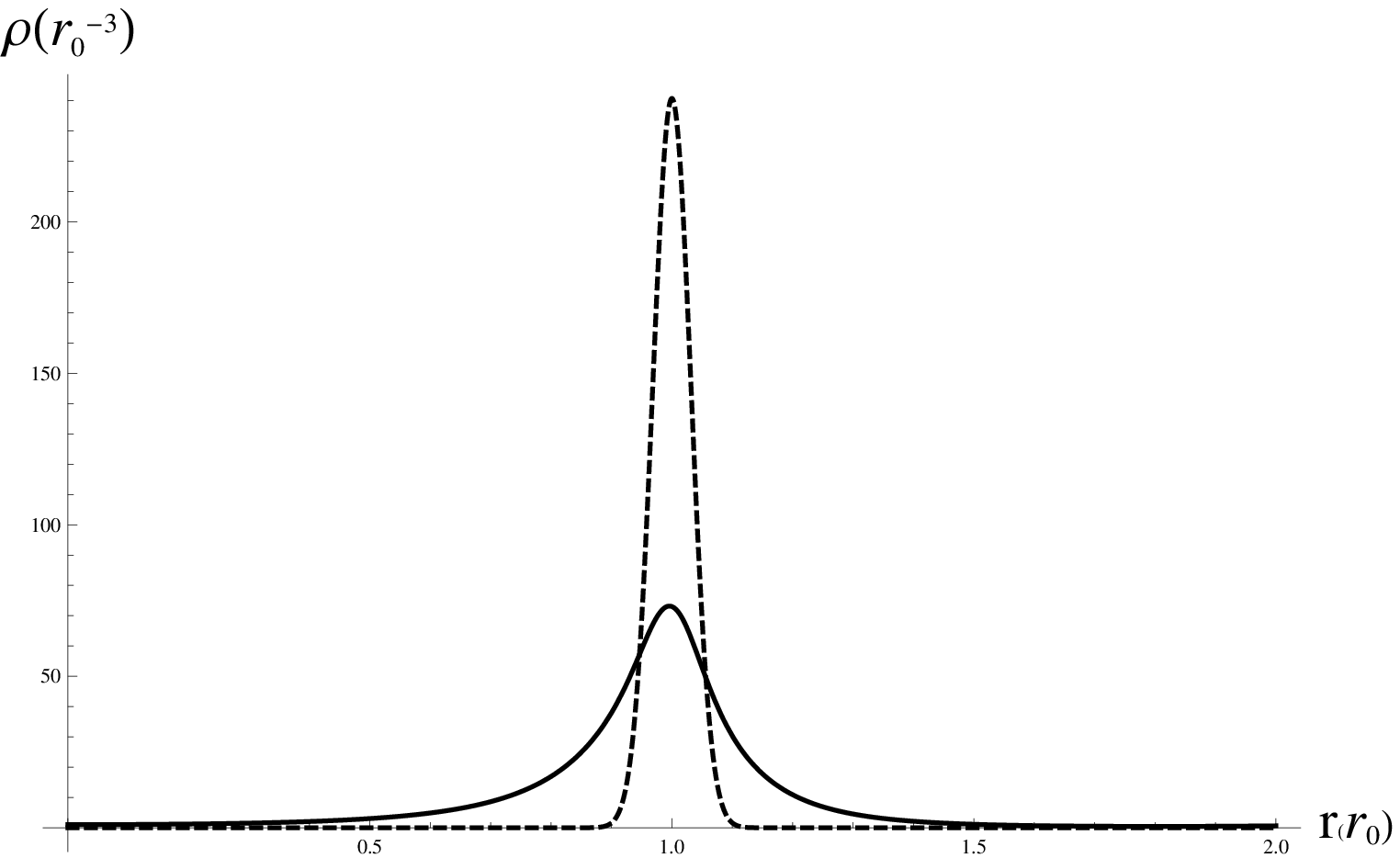}}\hspace{2mm} 
\subfloat{\includegraphics[width = 64mm]{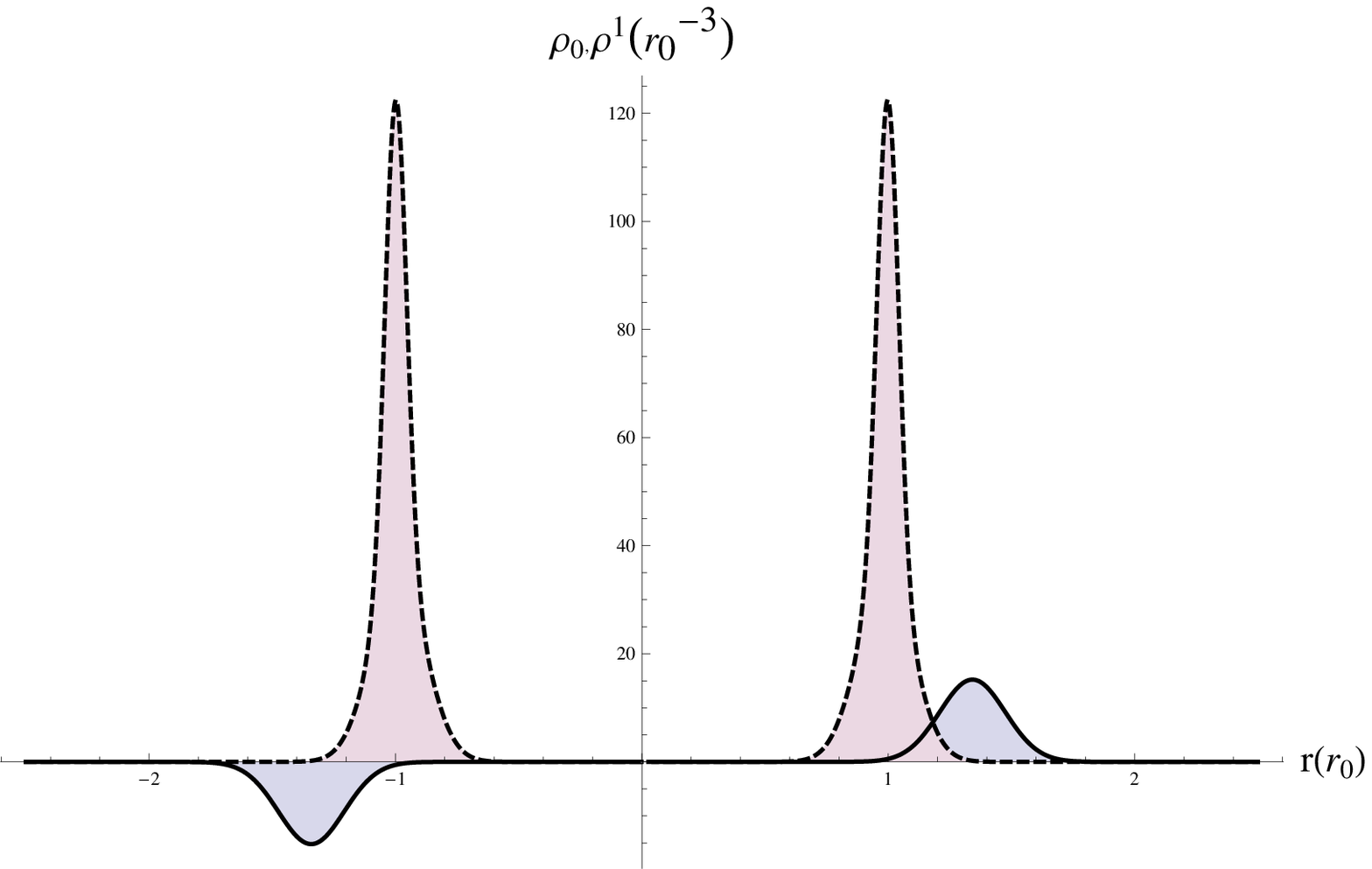}}\\
\subfloat{\includegraphics[width = 54mm]{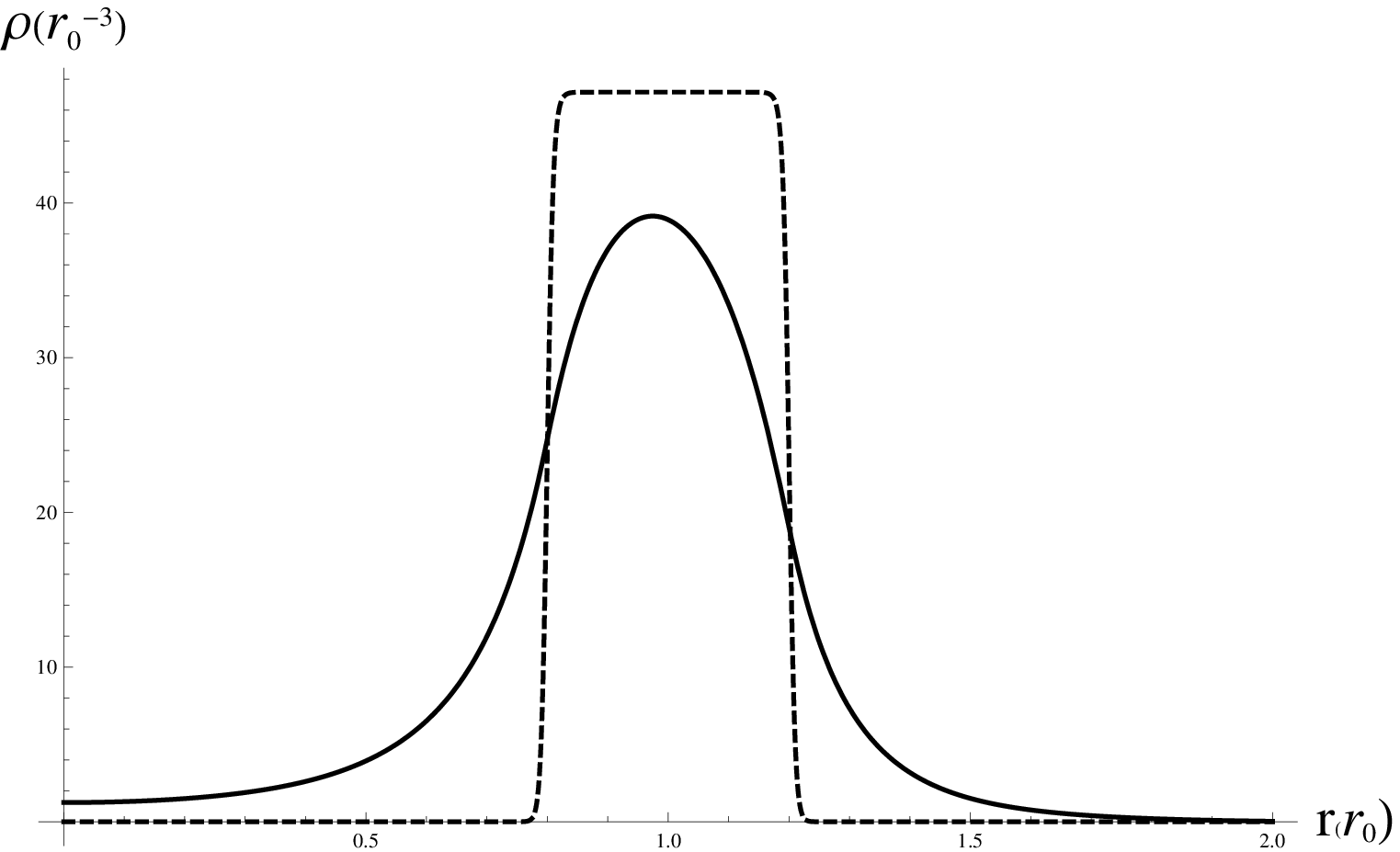}} \hspace{3mm}
\subfloat{\includegraphics[width = 64mm]{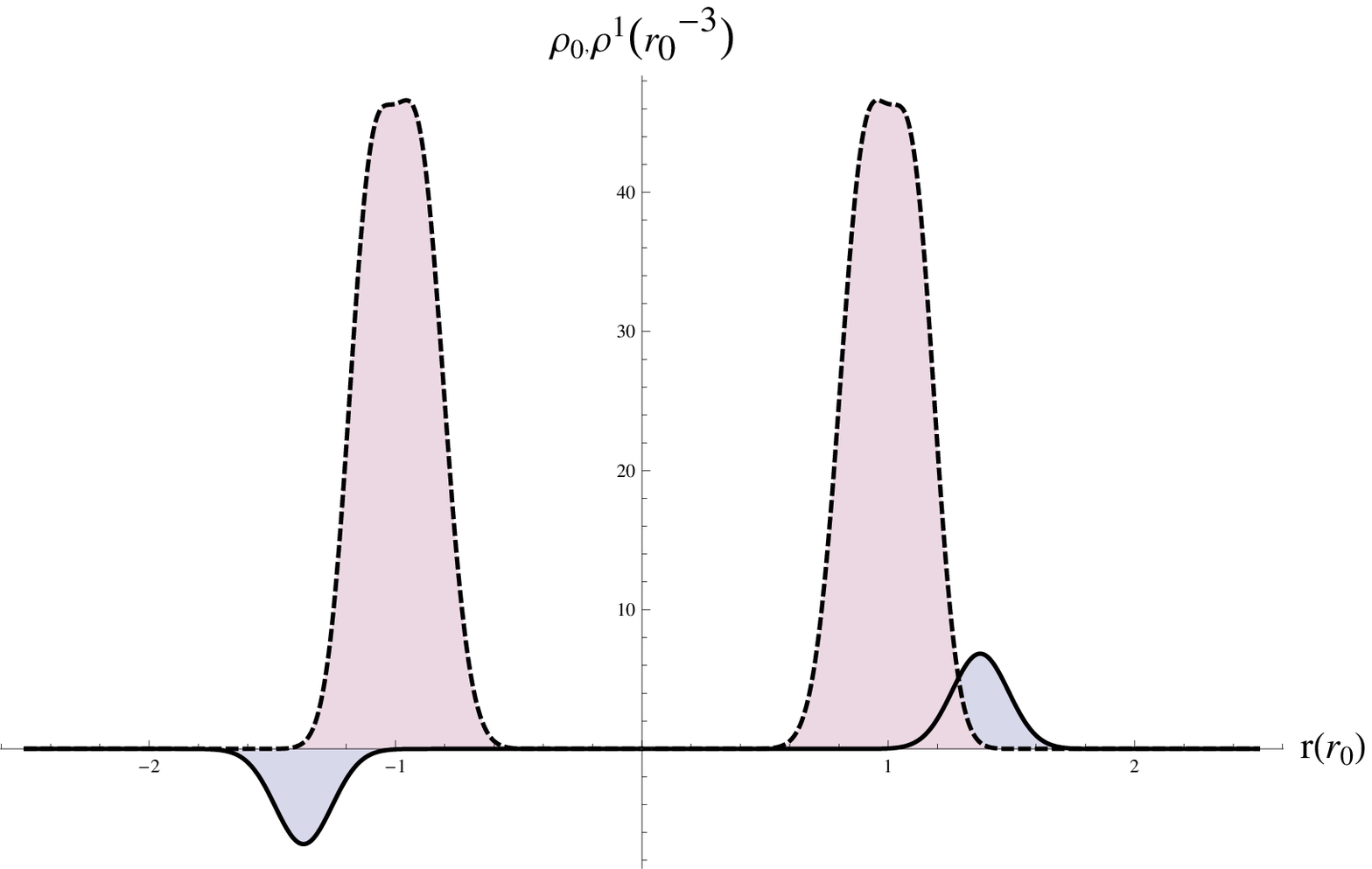}}
\caption{\footnotesize{a)Jellium density (Dashed line) and electron density (continuous line) $\rho$ for the $C60$ cluster in homogeneous shell parametrization;
b)) Ground-state electron density (dashed) and induced charge density (continuous) on the $Oz$ direction for homogeneous shell;
c))Jellium density (Dashed line) and electron density (continuous line) $\rho$ for the $C60$ cluster in narrow gaussian parametrization;
d) Ground-state electron density (dashed) and induced charge density (continuous) on the $Oz$ direction for narrow gaussian jellium}} 
\label{C60graf}
\end{figure}

The results in the electron density are physical and in good agreement with the experimental values for the inner, $\simeq 1.8 \mathring{A}$ and outer radius of the fullerene $\simeq 5.1 \mathring{A}$, see ~\ref{C60graf}a),c). From the calculations of dipolar polarizability, we have obtained the expected volumic shift in density on the direction of the potential gradient ($Oz$ axes) shown in Fig ~\ref{C60graf}b),d) while the value for polarizability is sweepings an interval between $80 \mathring{A}^3$ and $85 \mathring{A}^3$, depending on the chosen full width at half maximum of the gaussian jellium.

\begin{figure}[h]
\centering
\includegraphics[width=0.4\linewidth]{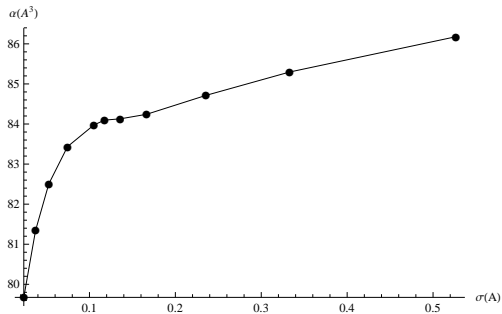}
\caption{Polarizability vs width of different parametrization for $C_{60}$ fullerene}
\label{fig:C60results}
\end{figure}

In Fig ~\ref{fig:C60results} we have plotted the results for $C_{60}$ polarizability for different parametrizations of the jellium model. The results are unexpectedly close to the experimental value of $78 A^3$. While the global aspect of ground state electron density has no essential dependence on $\Delta$, the values of the density far from center or the cluster influence the value of the polarizability, fact which explains the spectrum of obtained values.  

In the case of spherical homogeneous shell the obtained polarizability was $\alpha=92\mathring{A}^3$. The fact that the results from the gaussian parametrization of the jellium were closer to the experimental value raise the question of weather this aspect is a mathematical property of the equations involved, or simply the gaussian case is more realistic as discussed due to asymptotic tail of the core electrons from carbon atoms which must be taken into account in the geometry of jellium.

In order to test further the power of our equation and the validity of the approximations involved we calculate the polarizability of fullerenes with $180$ and $240$ carbon atoms also in a spherical symmetry and with a gaussian profile. The obtained results are around $260\mathring{A}^3$ for $C_{180}$ comparable with the RPA result of $300\mathring{A}^3$ \cite{zope2008static} and $340\mathring{A}^3$ for $C_{240}$ compared with $432\mathring{A}^3$ from RPA \cite{zope2008static}.

As the size of fullerene increases, the method starts to fail, one of the reasons being the fact that the spherical symmetry begins to be broken. Nonetheless, the results are still comparable with those from more involved methods \cite{gueorguiev2004quantum} from computational point of view.

\section*{Conclusions}

We exploit the Thomas-Fermi theory to compute the ground-state density of the electron system in a various number of \emph{Na} clusters and $C_{60}$ fullerene using the \emph{anzatz} of spherical symmetry and the jellium model for ionic background. Further the perturbation theory it is used to derive a differential equation in such TF systems for a general external one-body potential from which the induced change in the density of electrons can be derived and consequently the static linear response for any angular dependence or multipolarity.

This equation for multipolar moments it is solved for the same metallic clusters in the case of dipole external potential and the dipole polarizabilities are obtained. The errors are under $15 \%$ for the Sodium clusters while for fullerene, in a certain parametrization of the jellium model, we can obtain even the experimental value of the polarizability.

From all the semi-quantitative results, we conclude that our method is fast numerically and a good replacement for all the $ab$ $initio$ method which allow to compute the static linear response.

\end{document}